# Effect of Monolayer Thickness Fluctuations on Coherent Exciton Coupling in Single Quantum Wells

Yuri D. Glinka,[1,2] Zheng Sun,[1] Mikhail Erementchouk,[3] Michael N. Leuenberger,[3] Alan D. Bristow,[4,5] Steven T. Cundiff,[4] Allan S. Bracker[6] and Xiaoqin Li[1*]

[1]*Department of Physics, University of Texas, Austin, TX 78712, USA*
[2]*Institute of Physics, National Academy of Sciences of Ukraine, Kiev, 03028, Ukraine*
[3]*NanoScience Technology Center and Department of Physics, University of Central Florida, Orlando, FL 32826, USA*
[4]*JILA, University of Colorado and National Institute of Standard and Technology, Boulder, CO 80309-0440, USA*
[5]*Department of Physics, West Virginia University, Morgantown, WV 26506-6315, USA*
[6]*Naval Research Lab, Washington, DC 20375, USA*
*Email: elaineli@physics.utexas.edu

Monolayer fluctuations in the thickness of a semiconductor quantum well (QW) lead to three types of excitons, located in the narrower, average and thicker regions of the QW, which are clearly resolved in optical spectra. Whether or not these excitons are coherently coupled via Coulomb interaction is a long-standing debate. We demonstrate that different types of disorders in quantum wells distinctly affect the coherent coupling and that the coupling strength can be *quantitatively* measured using optical two-dimensional Fourier transform spectroscopy. We prove experimentally and theoretically that in narrow quantum wells the coherent coupling occurs predominantly between excitons residing in the disorder-free areas of QW's and those residing in plateau-type disorder. In contrast, excitons localized in fault-type disorder potentials do not coherently couple to other excitons.

Disorder at surfaces and interfaces plays an increasingly important role in ever-shrinking electronic devices. Even in nanostructures of the highest quality, monolayer fluctuations are inevitable. Importantly, monolayer fluctuations at interfaces of quantum wells (QWs) change the QW thickness, leading to wider or narrower regions that can be categorized as fault-type or plateau-type (Fig. 1(a)) disorder, respectively. Under optical excitation, bound electron-hole pairs (excitons) that form in such QWs are affected by the disorder potential and will often exhibit several spectrally distinct resonances [1-3], instead of a single inhomogeneously broadened resonance. Whether or not these different types of excitons are coherently coupled via Coulomb interactions is an outstanding and much debated question. Previous investigations have yielded conflicting results [4-8], partially due to the limited spectral and temporal information accessible using traditional spectroscopic methods and partially due to insufficient control of the disorder at the QW interfaces.

Understanding coherent interaction among multiple electronic states is a prerequisite to controlling material properties at the level of electrons, which is a ubiquitous challenge in material science. Specifically, the presence or absence of coherent coupling among spectrally resolved excitons significantly influence energy transfer [9], photon emission statistics [10], and even quantum-logic operations [11] in semiconductor heterostructures such as QWs, quantum wires, and quantum dots. This problem is also relevant for a broader range of materials. For example, exciton dissociation in conjugated polymers occurs in a two-step process: the formation of a charge-transfer state at an energy level above the initial exciton followed by a complete dissociation process [12, 13]. The formation of the charge-transfer state is dictated by the interplay of the disorder and the Coulomb interaction, which is the essence to the problem of interest here.

In this paper, we investigate coherent coupling between different types of excitons in a single, narrow GaAs/AlGaAs QW using the powerful technique of optical two-dimensional Fourier transform spectroscopy (2DFTS) [14-19]. We first identify three types of excitons confined in different regions of the QW. We refer to the excitons confined in the wider, average-thickness, and narrower regions of the QW as type A, B, and C excitons, respectively, as illustrated in Fig. 1(a). Using 2DFTS, we are able to separate complex quantum-mechanical pathways in the coherent nonlinear response and unambiguously identify the presence of coherent coupling between type B and C excitons. In contrast, such coupling is missing between type A and B excitons. This difference in exciton coupling originates from the nature of exciton resonances: type A excitons are bound states mainly localized within the fault-type disorder potential; type B are delocalized in the perfect QW region (or disorder-free regions); and type C excitons are scattering resonances associated with plateau-type defects. Different spatial overlap between the excitons is responsible for the absence or presence of coupling in the 2D spectra. This conclusion is supported by calculations based on a single-defect model.

2DFTS is a heterodyne-detected four-wave mixing (FWM) technique, which monitors and correlates nonlinear polarization phase evolution during two independent periods, $\tau$ and $t$, separated by a waiting period $T$, as illustrated in Fig. 1(b). A 2D spectrum as a function of the absorption frequency $\omega_\tau$ and emission frequency $\omega_t$ is then obtained by Fourier transforming the FWM signal with respect to time variables $\tau$ and $t$. A diagonal peak in a 2D spectrum indicates that an oscillation at absorption frequency $\omega_\tau$ during the first time period gives rise to an oscillation at emission frequency $\omega_t$ during the third period. Coupling between resonances can be identified by the presence of cross-peaks in 2D spectra, for which $\omega_\tau \neq \omega_t$. Therefore, 2DFTS is particularly suitable for quantifying coupling among electronic transitions, as quantum-mechanical pathways associated with coupling are isolated in the spectra.

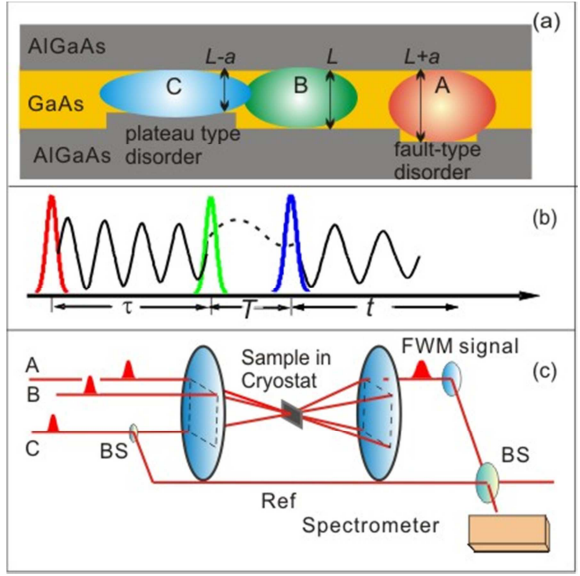

Fig. 1 (color online): (a) illustration of different types of excitons form in a disordered QW. (b) Pulse sequence and (c) experimental set-up in 2DFTS.

The experimental setup is described in detail elsewhere [20]. Briefly, three phase-stabilized, collinearly polarized excitation pulses with wave vectors $\boldsymbol{k}_a$, $\boldsymbol{k}_b$, and $\boldsymbol{k}_c$ are arranged in box geometry generating a complex FWM signal in the phase-matched direction: $\boldsymbol{k}_s = -\boldsymbol{k}_a + \boldsymbol{k}_b + \boldsymbol{k}_c$, as illustrated in Fig. 1(c). The excitation pulse sequence chosen is the rephasing time ordering, where the conjugate pulse ($\boldsymbol{k}_a$) arrives first and $\boldsymbol{k}_c$ arrives last. The evolution time ($\tau$) between $\boldsymbol{k}_a$ and $\boldsymbol{k}_b$ pulses, the waiting time ($T$) between pulses $\boldsymbol{k}_b$ and $\boldsymbol{k}_c$, and the emission time ($t$) after the arrival of pulse $\boldsymbol{k}_c$ govern the complex rephased signal, $S(\tau, T, t)$. We resolve the phase information of the signal field, which is made possible by heterodyne detection with a phase-stabilized reference beam and stepping the excitation pulse delays with interferometric precision. The emission frequency ($\omega_t$) is determined by sending the signal through a spectrometer and the absorption frequency ($\omega_\tau$) is retrieved by a numerical Fourier transform of the $S(\tau, T, \omega_t)$ data with respect to $\tau$.

We studied a series of GaAs/Al$_{0.3}$Ga$_{0.7}$As single QW's with four different thickness grown by molecular beam epitaxy on a GaAs(100) substrate. The nominal thicknesses are 4.2, 6.2, 8.4, and 14.0 nm, respectively. Two-minute growth interruption at the interfaces introduces monolayer-width fluctuations with relatively large lateral dimensions on the order of tens of nanometers. The sample was held at 4.2 K. The average laser excitation power was between 3 to 5 mW corresponding to an estimated sheet exciton density ~ 2.0 - 5.8×10$^{10}$ cm$^{-2}$. At this high excitation power, one expects excitation-induced dephasing [21, 22] to contribute significantly to the line width of the exciton resonance.

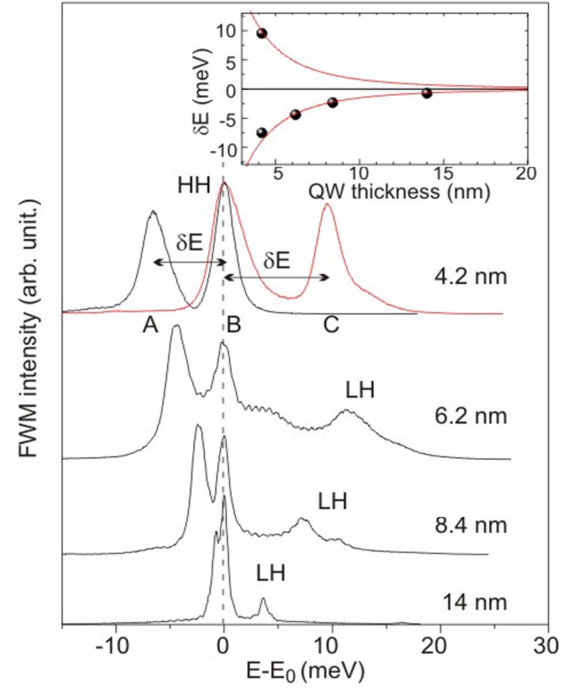

Fig. 2 (color online): FWM spectra plotted with relative photon energy ($E - E_0$), where $E_0$ specifies the position of HH excitons residing in the regions of the perfect QW. Two spectra from the 4.2 nm QW are taken for different laser tuning. Inset: QW width dependence of the HH exciton resonance splitting due to the monolayer thickness fluctuations at the interface (dots) relative to B-type excitons. The red lines present the result of theoretical fits.

We first identify relevant exciton resonances via FWM spectra displayed in Fig. 2. The heavy-hole (HH) and light-hole (LH) excitons are split by confinement in the growth direction in a QW. The LH excitons are clearly observed and labeled for QWs with nominal thickness of 6.2, 8.4, and 14 nm. The LH exciton in the 4.2 nm QW is shifted further to the higher energy, outside the excitation bandwidth. We will focus on HH resonances only in the current study.

Close inspections reveal that HH resonances are split into two or three resonances, which arise from the monolayer fluctuations of the QW thickness. This assignment can be proven by investigating the systematic change of the monolayer splitting as a function of the average QW width. We initially make the reasonable assumption that the lowest energy peak originates from type A excitons residing in the fault-type disorder potential. This allows us to plot the FWM spectra as a function of the relative energy $\delta E = (E - E_0)$, where $E_0$ specifies the energy for type B excitons in the perfect QW. Negative $\delta E$ corresponds to type A excitons residing in fault-type disorders, while positive $\delta E$ corresponds to type C excitons formed near plateau-type disorder. We then plot $\delta E$ as a function of the QW thickness in the inset of Fig. 2. Within the effective mass approximation, $\delta E$ due to monolayer fluctuation of the QW width ($\delta L^* = \pm a$) is

$$\delta E = \frac{\hbar^2 \pi^2}{2\mu_{HH}} \left( \frac{1}{(L^* \pm a)^2} - \frac{1}{L^{*2}} \right) \approx -\frac{\hbar^2 \pi^2 \delta L^*}{\mu_{HH} L^{*3}}, \quad (1)$$

where $L^* = L + 2\Lambda$ denotes an enlargement of the QW width owing to an average wave function penetration depth into the $Al_{0.3}Ga_{0.7}As$ barrier ($\Lambda = 1.5$ nm) [23]. We first extracted the in-plane reduced exciton mass $\mu_{HH} = 0.055\,m_e$ from the dependence of $E_0$ on the QW width (not shown explicitly). We then fitted the data in Fig. 2 using Eq. (1) to obtain $a = 0.25$ nm, which matches the known value for the thickness of one atomic monolayer in GaAs. This analysis supports our assignment of different types of excitons. We focus the rest of the paper on 2DFTS experiments performed on the 4.2 nm thick QW as it clearly exhibited all three types of HH excitons [24].

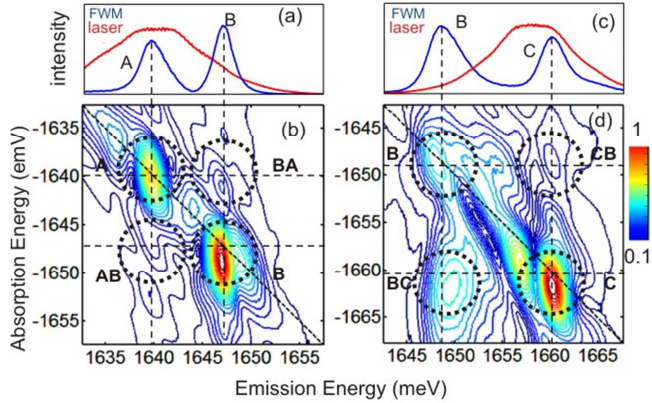

Fig. 3 (color online): FWM and 2DFTS amplitude spectra taken on the 4.2 nm QW at $T = 125\,fs$ with co-linearly polarized excitation beams (a-b) when exciton A and B resonances are excited and (c-d) when exciton B and C resonances are excited.

The laser bandwidth is ~ 16 meV; hence, two 2D spectra were acquired with the laser tuned to optimally excite two of the three HH exciton resonances at one time. The waiting time was set to be $T = 125\,fs$ to avoid temporal pulse overlap. Figure 3 shows the measured 2D amplitude spectra together with the corresponding regular FWM and laser spectra. We did not observe clear evidence of coherent coupling between exciton A and B in Fig. 3(b). In contrast, coherent coupling between exciton resonance B and C is clearly identified as the cross-peak BC in Fig. 3(d). One may expect another cross-peak at the location indicated by CB. Previous work has documented that 2D spectra of semiconductor QWs display a strong asymmetry between cross peaks[25], which explains the absence of a cross peak at location indicated by CB.

Prior experiments aimed at investigating coherent coupling among different types of excitons only permitted one to look for qualitative signatures, such as oscillatory behavior in time- or spectrally resolved FWM experiments [4, 5]. The interpretation of these experiments is challenging, due to the limited spectral or temporal information accessible to conventional spectroscopy. In contrast, 2DFTS allows us to quantify the strength of coherent coupling between these resonances, rather than only asking whether such coupling exists. In the case of a V-level system, for example, a strong coherent coupling induced by a common electronic state should lead to the maximum cross-peak intensity to be the geometric average of the two diagonal peak intensities [26]. In Fig. 3(d), the coupling between resonances B and C is quantified by the ratio between the cross-peak BC and the geometric average of the diagonal peak intensities, giving a value of ~ 0.6 that is significantly above the noise level. In contrast, we observe no clear cross-peaks in Fig. 3(b). We place an upper bound for the coupling between resonances A and B to be below 0.1, limited by the noise level in the 2D spectra. The strong cross-peak intensity BC at the chosen $T$ suggests that it arises from electronic coherent coupling and not population relaxation. We confirm that the relaxation lifetimes of the excitons are beyond tens of picoseconds with spectrally and temporally resolved pump-probe measurements (data not shown). The slow relaxation process cannot account for the strong cross-peak BC in the 2D spectra (Fig. 3(d)) at the short waiting time. Incoherent coupling due to population relaxation between excitons confined in different thickness regions of a QW has been investigated previously using 2DFTS and observed at waiting times at 20 ps or longer [27]. Another recent experiment conducted on individual excitons localized in a QW presented evidence of coherent coupling in certain regions of the sample [8], possibly consistent with our experimental findings.

We briefly comment on other features of the 2D spectra, which are not essential for identifying the coupling or quantifying the coupling strength. There is a peak near the diagonal peak C in Fig. 3(d), which may be due to the formation of unbound biexcitons investigated in other recent 2DFTS experiments [28-30]. There are small satellite peaks near the diagonal line that arise from the direct scattering of excitation pulses [20]. These spurious peaks do not contaminate the measurements of the cross-peaks. Thus, no error was introduced in the quantitative evaluation of the coupling strength. Finally, the lineshape of the peaks (especially the diagonal peaks) are elongated along the absorption frequency axis. We attribute the elongation to the relatively high laser power used in the experiments and it may be related to the excitation-induced dephasing.

To support the interpretation of the measured 2D spectra, we performed calculations based on microscopic theory of semiconductor coherent nonlinear response in QWs with in-plane inhomogeneous disorder potentials [31]. Within the $\chi^{(3)}$-approximation [32-35] the equation of motion for the nonlinear polarization [36, 37] taking into account the HH excitons only is given by

$$\left( i\frac{\partial}{\partial t} + i\gamma - \hat{H}(\mathbf{R}) \right) P^{(3)}(\mathbf{R}, t) = F^{(3)}(\mathbf{R}, t). \quad (2)$$

To reproduce the qualitative features of 2D spectra, we can neglect the effect of Pauli-blocking and invoke the short-memory approximation[31, 38] to account for the Coulomb correlation. Under these approximations, the driving term is

$F^{(3)}(\mathbf{R}, t) = -\frac{1}{2} \int d\mathbf{R}_{1,2,3}\, \beta(\mathbf{R}, \mathbf{R}_1, \mathbf{R}_2, \mathbf{R}_3)\, p^*(\mathbf{R}_1, t) p(\mathbf{R}_2, t) p(\mathbf{R}_3, t),$ (3)
which includes an effective four-point potential $\beta$ to describe the Coulomb interaction and the linear optical response $p(\mathbf{R}, t)$. The latter is governed by an equation of the same form as Eq. (2) but with the source term $F^{(1)}(\mathbf{R}, t) = \vec{d} \cdot \vec{E}(\mathbf{R}, t)$, where $\vec{d}$ is the dipole moment and $\vec{E}$ is the electric field of the excitation. Taking the energy of the HH exciton in the perfect QW as zero, the Hamiltonian in Eq. (2) can be written as $\widehat{H}(\mathbf{R}) = -\frac{1}{2M}\nabla_{\mathbf{R}}^2 + W(\mathbf{R})$, where $M$ is the exciton effective mass, and $W(\mathbf{R})$ describes the random potential due to interface fluctuations [1, 39].

To emphasize the difference between excitons of different types, we adopt two additional approximations. First, we take into account that the effective potential $\beta$ in Eq. (3) decays fast with distance ($\propto 1/r^6$) and, thus, we treat it as a contact interaction. Second, we employ the fact that experimentally obtained spectra have well resolved resonances and retain only the main resonant contributions to the 2D spectrum. Using these assumptions, we find that the peaks in the 2D amplitude spectrum are given by
$P_{FWM}^{(3)}(\omega_t, \omega_\tau) \sim \sum_{i,j} \frac{\alpha_{i,j}}{(\omega_\tau + \omega_i - i\gamma_i)(\omega_t - \omega_j + i\gamma_j)(\omega_t - \omega_j + 2i\gamma_i + i\gamma_j)}$,
where $i$ and $j$ are the indices of the three excitons $\{A, B, C\}$, the overlap factors are $\alpha_{i,j} \sim \frac{1}{S}\int d\mathbf{R}\, |p(\mathbf{R}, \omega_i)|^2 |p(\mathbf{R}, \omega_j)|^2$, and $S$ is the area of the QW. Thus the amplitude of a peak in the 2D spectra, within described approximations, is determined by the spatial overlap of respective linear response functions.

Exciton linear response functions $p(\mathbf{R}, \omega_i)$ vary significantly depending on the type of excitons. To illustrate this point, we performed a simple calculation of linear response near a single circular defect in an otherwise perfect QW. The calculated linear response near a fault- and a plateau-type defect at corresponding resonant frequencies are shown in Fig. 4(a) and 4(b), respectively. Type A excitons (solid red line in Fig. 4(a)) correspond to bound states that are localized within fault-type defects and decay exponentially outside. This fast spatial decay leads to small overlap with type $B$ excitons (dashed green curves in Fig. 4(a) and 4(b)) and even less overlap with type $C$ excitons (solid blue curve in Fig. 4(b)). In contrast, type $C$ excitons are unbounded scattering resonances that decay algebraically outside of plateau-type defects. Type B excitons largely occupy the perfect QW region, but their linear response functions are perturbed near either type of defects. Their penetration into the defects depends on whether it is a fault- or plateau-type defect. The strength of the cross-peaks is determined by $\alpha_{i,j}$, which is a two-dimensional spatial integral over the linear response $p(\mathbf{R}, \omega)$, at different resonance frequencies $\omega_{i,j}$. Qualitatively, the large spatial overlap between excitons B and C leads to coherent coupling while the minimal overlap between excitons A and B accounts for the lack of coupling as illustrated by Fig. 4(a) and 4(b).

Simulated 2D spectra are shown in Fig. 4(c) and 4(d), where the latter confirm coherent coupling between exciton types B and C. The calculations neglected the incoherent population relaxation dynamics and included several parameters including dephasing times and overlap factors ($\alpha_{AB}$, $\alpha_{BC}$ and $\alpha_{AC}$) determined from the single circular defect calculations. Use of the single-defect model is supported by the low spatial density of monolayer disorders evidenced by photoluminescence measurements (data now shown). Although the simulated spectra cannot make a quantitative prediction of coupling strength due to the lack of detailed information on the disorder potential, the qualitative features on the presence and absence of coherent coupling are reproduced robustly.

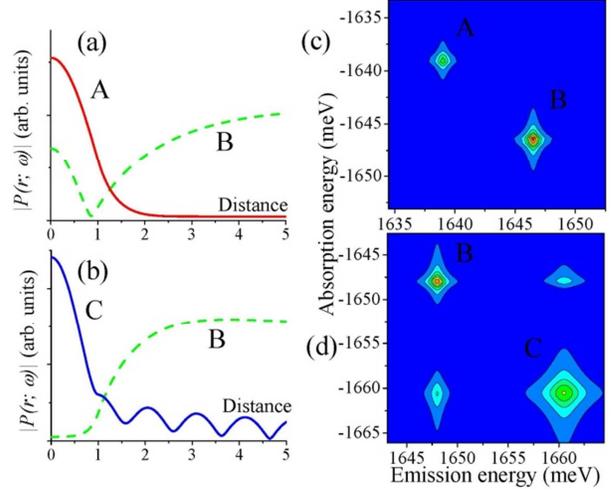

Fig. 4 (color online): Calculated linear response function vs. distance (normalized to the radius of the defect) to the center of a single circular defect for (a) fault-type disorder and (b) plateau-type disorder. Green dashed lines are for type B excitons. Solid red line is for type A excitons in (a), and solid blue line is for type C excitons in (b). Simulated 2D spectra when (c) type A and type B excitons are excited and (d) type B and type C excitons are excited.

In summary, the answer to whether or not exciton resonances localized in different regions of a disordered QW are coupled depends on the nature of the exciton resonances. Electronic 2DFTS provides unambiguous evidence of coupling and quantitative measurement of the coherence coupling strength. The theoretical and experimental approach and findings reported here are applicable to a broad range of problems including energy transfer in natural/artificial photosynthetic systems [40-42] and charge transfer in conjugated polymers.

Acknowledgement: Authors at UT-Austin gratefully acknowledge financial support from the following sources: NSF DMR-0747822, Welch Foundation F-1662, ARO-W911NF-08-1-0348, and the Alfred P. Sloan Foundation. M.N.L. acknowledges support from NSF (Grant ECCS-0901784), AFOSR (Grant FA9550-09-1-0450), and NSF (Grant ECCS-1128597).

**References**

[1] H. Castella and J. W. Wilkins, Phys. Rev. B **58**, 16186 (1998).
[2] V. Savona and W. Langbein, Phys. Rev. B **74**, 075311 (2006).
[3] D. Bimberg, J. Christen, T. Fukunaga, H. Nakashima, D. E. Mars, and J. N. Miller, J. Vac. Sci. Technol. B **5**, 1191 (1987).
[4] E. O. Göbel, K. Leo, T. C. Damen, J. Shah, S. Schmitt-Rink, W. Schäfer, J. F. Müller, and K. Köhler, Phys. Rev. Lett. **64**, 1801 (1990).
[5] M. Koch, J. Feldmann, E. O. Göbel, P. Thomas, J. Shah, and K. Kohler, Phys. Rev. B **48**, 11480 (1993).
[6] A. Euteneuer, et al., Phys. Rev. Lett. **83**, 2073 (1999).
[7] M. Phillips and H. L. Wang, solid state commun. **111**, 317 (1999).
[8] J. Kasprzak, B. Patton, V. Savona, and W. Langbein, Nat. Photonics **5**, 57 (2011).
[9] T. Takagahara, Phys. Rev. B **31**, 6552 (1985).
[10] R. Singh and G. Bester, Phys. Rev. Lett. **103**, 063601 (2009).
[11] P. C. Chen, C. Piermarocchi, L. J. Sham, D. Gammon, and D. G. Steel, Phys. Rev. B **69** (2004).
[12] M. Pope and C. E. Swenberg, *Electronic Processes in Organic Crystals and Polymers* (Oxford University Press, New York, 1999).
[13] C. L. Braun, J. Chem. Phys. **80**, 4157 (1984).
[14] X. Li, T. Zhang, C. N. Borca, and S. T. Cundiff, Phys. Rev. Lett. **96**, 057406 (2006).
[15] S. Mukamel, Annu. Rev. Phys. Chem **51**, 691 (2000).
[16] M. Cho, Chem. Rev. **108**, 1331 (2008).
[17] I. Kuznetsova, N. Gogh, J. Forstner, T. Meier, S. T. Cundiff, I. Varga, and P. Thomas, Phys. Rev. B **81**, 075307 (2010).
[18] A. D. Bristow, T. H. Zhang, M. E. Siemens, S. T. Cundiff, and R. P. Mirin, J. Phys. Chem. B **115**, 5365 (2011).
[19] J. A. Davis, C. R. Hall, L. V. Dao, K. A. Nugent, H. M. Quiney, H. H. Tan, and C. Jagadish, J. Chem. Phys. **135** (2011).
[20] A. D. Bristow, D. Karaiskaj, X. Dai, T. Zhang, C. Carlsson, K. R. Hagen, R. Jimenez, and S. T. Cundiff, Rev. Sci. Instrum. **80**, 073108 (2009).
[21] H. Wang, K. Ferrio, D. G. Steel, Y. Z. Hu, R. Binder, and S. W. Koch, Phys. Rev. Lett. **71**, 1261 (1993).
[22] Y. Z. Hu, R. Binder, S. W. Koch, S. T. Cundiff, H. Wang, and D. G. Steel, Phys. Rev. B **49**, 14382 (1994).
[23] P. V. Santos, M. Willatzen, M. Cardona, and A. Cantarero, Phys. Rev. B **51**, 5121 (1995).
[24] We performed 2DFTS experiments on all QWs and observed systematic changes in the cross-peaks. For example, no coupling was observed between any HH excitons in the widest QW. Coupling between HH-LH in differet regions was observed in the QW with the intermediate thickness. It is outside the scope of the current paper to present results from all QWs.
[25] C. N. Borca, T. H. Zhang, X. Q. Li, and S. T. Cundiff, Chem. Phys. Lett. **416**, 311 (2005).
[26] This statement is true under the condition that both dipole transitions are excited by equal electric field strength.
[27] G. Moody, M. E. Siemens, A. D. Bristow, X. Dai, A. S. Bracker, D. Gammon, and S. T. Cundiff, Phys. Rev. B **83**, 245316 (2011).
[28] K. W. Stone, K. Gundogdu, D. B. Turner, X. Q. Li, S. T. Cundiff, and K. A. Nelson, Science **324**, 1169 (2009).
[29] D. Karaiskaj, A. D. Bristow, L. J. Yang, X. C. Dai, R. P. Mirin, S. Mukamel, and S. T. Cundiff, Phys. Rev. Lett. **104**, 117401 (2010).
[30] L. Yang and S. Mukamel, Phys. Rev. Lett. **100**, 057402 (2008).
[31] T. Östreich, K. Schönhammer, and L. J. Sham, Phys. Rev. B **58**, 12920 (1998).
[32] M. Lindberg, Y. Z. Hu, R. Binder, and S. W. Koch, Phys. Rev. B **50**, 18060 (1994).
[33] K. Victor, V. M. Axt, and A. Stahl, Phys. Rev. B **51**, 14164 (1995).
[34] W. Schäfer, D. S. Kim, J. Shah, T. C. Damen, J. E. Cunningham, K. W. Goossen, L. N. Pfeiffer, and K. Köhler, Phys. Rev. B **53**, 16429 (1996).
[35] N. H. Kwong, R. Takayama, I. Rumyantsev, M. Kuwata-Gonokami, and R. Binder, Phys. Rev. B **64**, 045316 (2001).
[36] M. Erementchouk, M. N. Leuenberger, and L. J. Sham, Phys. Rev. B **76**, 115307 (2007).
[37] M. Erementchouk, M. N. Leuenberger, and X. Q. Li, Physica Status Solidi C **8**, 1141 (2011).
[38] S. Savasta, O. Di Stefano, and R. Girlanda, Phys. Rev. Lett. **90**, 096403 (2003).
[39] R. Zimmermann and E. Runge, Phys. Status Solidi A-Appl. Res. **164**, 511 (1997).
[40] G. S. Engel, T. R. Calhoun, E. L. Read, T. K. Ahn, T. Mancal, Y. C. Cheng, R. E. Blankenship, and G. R. Fleming, Nature **446**, 782 (2007).
[41] A. Ishizaki and G. R. Fleming, J. Phys. Chem. B **115**, 6227 (2011).
[42] E. Collini, C. Y. Wong, K. E. Wilk, P. M. G. Curmi, P. Brumer, and G. D. Scholes, Nature **463**, 644 (2010).